\documentclass[aps,prl,twocolumn,floatfix]{revtex4-1}

\usepackage{bm,bbm,amsmath, amssymb,amsbsy, amsthm,graphicx,subfigure,color,txfonts,multirow,relsize}
\usepackage[framemethod=tikz]{mdframed}
\usepackage{mathtools}
\usepackage[colorlinks]{hyperref}
\hypersetup{
	bookmarksnumbered,
	pdfstartview={FitH},
	citecolor={blue},
	linkcolor={red},
	urlcolor={black},
	pdfpagemode={UseOutlines}
}

 \newcommand{\tr}[1]{\text{Tr}}
\newcommand{\ket}[1]{|#1\rangle}
\newcommand{\bra}[1]{\langle#1|}

\begin{document}
 \title{An axiomatic measure of one-way quantum information}

\author{Davide Girolami}
\email{davegirolami@gmail.com}
\affiliation{$\hbox{Los Alamos National Laboratory, Theoretical Division, Los Alamos, NM, 87545, USA}$
 }
 \affiliation{$\hbox{Politecnico di Torino, Corso Duca degli Abruzzi 24, Torino, 10129, Italy}$
 }

\begin{abstract}
I introduce an algorithm to detect one-way quantum information between two interacting quantum systems, i.e. the direction and orientation of the information transfer in arbitrary quantum dynamics. 
  I then build  an information-theoretic quantifier of one-way information which satisfies a set of desirable axioms.  In particular, it correctly evaluates whether correlation implies one-way quantum information, and when the latter is transferred between uncorrelated systems.
   In the classical scenario,  the quantity measures information transfer between random variables. I also generalize the method to identify and rank concurrent sources of quantum information flow in many-body dynamics,  enabling to reconstruct causal patterns in complex networks.  
 \\
 \ \\
 Keywords: quantum information, quantum correlations, quantum dynamics.
  
\end{abstract}

\date{\today}
 
\maketitle
\section{Introduction} 
 \noindent One-way quantum information manifests when the output state of a system in a process  is determined by its interaction with another system, but not {\it vice versa}. One-way information transfer can be associated to causal relations. A vast literature has discussed the problem of inferring causation from data in both classical and quantum scenarios \cite{granger,massey,pearl,shalizi,rubin,james,janzing,tucci}, because of its importance   for Science. 
Yet, a crucial problem is still unsolved: how can we {\it quantify} one-way information between quantum  systems?  In general, there is no consensus about how to measure  the peculiar one-way information flow that characterizes causation.  Given the state of a quantum system, measures of  quantum correlations mark well the amount of information shared by the components of the system in terms of entropic or geometric quantifiers \cite{modirev,horo}. However, given a multipartite quantum channel, we do not have any reliable metric to evaluate the information transferred during its implementation.  Unfortunately, widely employed causation measures misinterpret causal links between classical random variables in simple case studies \cite{janzing,james}, so we cannot just translate them in the quantum regime.\\
  

\noindent Here, I construct an information-theoretic measure of one-way information (OWI), capturing the direction of the   information flow between causally connected systems.  OWI is  exemplified by a measuring probe that updates its state based on the information acquired from a measured system. A controlled gate is then an adequate mathematical characterization for OWI flow from a system to another.  Another example of OWI is the instruction that a controlling device sends to regulate the state of a controlled machine.\\

\noindent First, I focus on the problem of inferring OWI in an arbitrary quantum channel. I present a three-step algorithm which discovers and evaluates OWI given the input/output states of many-body quantum processes. In other words, it can  discriminate  different causal relations from same-looking input/output data. Also, it is experimentally implementable with current technology.  The scheme builds on previous proposals for evaluating causation \cite{chiri,costa,chirio,prx,chaves,leifer,modicaus,temp,cava,shap}, which yet did not fully address the problem of quantifying OWI by provably rigorous measures.  \\

\noindent Then, I build the OWI quantifier, which is calculated in the output state of the algorithm. I show that the quantity meets a set of  important properties, which are not satisfied by widely employed measures in classical information theory.  Specifically,  it vanishes when there is no information transfer. Unlike correlation quantifiers,   it unambiguously pinpoints the source and the recipient of the information.   It reliably describes the interplay between correlation and causation, capturing when correlation does imply causation, and when causation exists without correlation. In the classical scenario, it quantifies the amount of information transferred between random variables.   
  Notably, I show that when the algorithm is run by a quantum computer  \cite{nielsen}, even one of the currently available toy models, can evaluate OWI between classical systems that are untraceable by a classical device which implements an equivalent scheme.   Finally, the method is extended to quantify OWI in multipartite systems. I build a  measure of conditional causation that satisfies two important properties. First, it localizes the source of information, i.e. the measured system(s), in three or more interacting parties. Second, it ranks   multiple concurrent sources in terms of how much they affect, i.e. control, the evolution of a  target system. Consequently, it makes possible to quantitatively describe causal patterns in many-body dynamics. \\

\section{Quantifier of OWI}
 An instance of OWI is the   coupling of an apparatus $B$ with a measured system $A$. The interaction is formalized as  a controlled operation $C^{i\rightarrow j}_{A\rightarrow B}\sum_i c_i \ket{i j}_{AB} = \sum_i c_i \ket{i j\oplus i}_{AB}, \{i,j=0,1,\ldots, d-1\}$.  Indeed, the controlled gate is the logic operation related to the pre-measurement step in the ubiquitous Von Neumann meaasurement scheme \cite{vn}. Here, $\log_2d$ bits of information flows from $A$ to $B$. Consider now the evolution of   a bipartite quantum system $AB$ initially prepared in the state $\rho_A\otimes\rho_B$, which is described by the unitary transformation  
    $\rho^{U}_{AB}=U_{AB}\rho_{A}\otimes \rho_{B}U_{AB}^{\dagger}$.  
Assuming that the channel $U$ is unknown,  the goal is to quantify how much the dynamics of system $A(B)$  influences the dynamics of $B(A)$. 
 The question to answer is then: ``How much information $A(B)$ transfers to $B(A)$ via the channel $U$?''  The task is hard because, given the same  initial  state, different causal relations can produce the same output. I list  two pieces of evidence.\\
First, the roles of  control and target systems are basis-dependent.  For example, the two-qubit  CNOT  is equal to a controlled gate with swapped control and target qubits, and different measurement basis,  $C^{0,1\rightarrow 0,1}_{A\rightarrow B}={C}^{+,-\rightarrow +,-}_{B\rightarrow A} , \ket{\pm}=(\ket{0}\pm\ket{1})/\sqrt2$ \cite{smolin,rieffel}. The system $A$ therefore exerts maximal influence on $B$ via the controlled operation   $C^{i\rightarrow j}_{A\rightarrow B}$  only with respect to the bases $\{i\},\{j\}$.
This also implies that calculating the correlations in the input/output states, or the ability of  a channel to create correlations \cite{zanardi,luo2}, is  insufficient for drawing conclusions on  the information flow from a causing device to an affected system.
 When the roles of control and target system are inverted, the information travels in the opposite direction, while creating the same amount of  correlations.\\
Second, there can be causal links with neither initial nor final correlations. For example, $\text{C}^{0,1\rightarrow 0,1}_{A\rightarrow B} \ket{10}_{AB}= \ket{11}_{AB}$ is a causal relation, conversely to the local bit flip $X_B\ket{10}_{AB}=\ket{11}_{AB}$. They are two different physical processes that generate the same output from the same initial state \cite{note}.\\
   \begin{figure}[t!]
\hspace{-5pt}
\includegraphics[width=.48\textwidth,height=3.7cm]{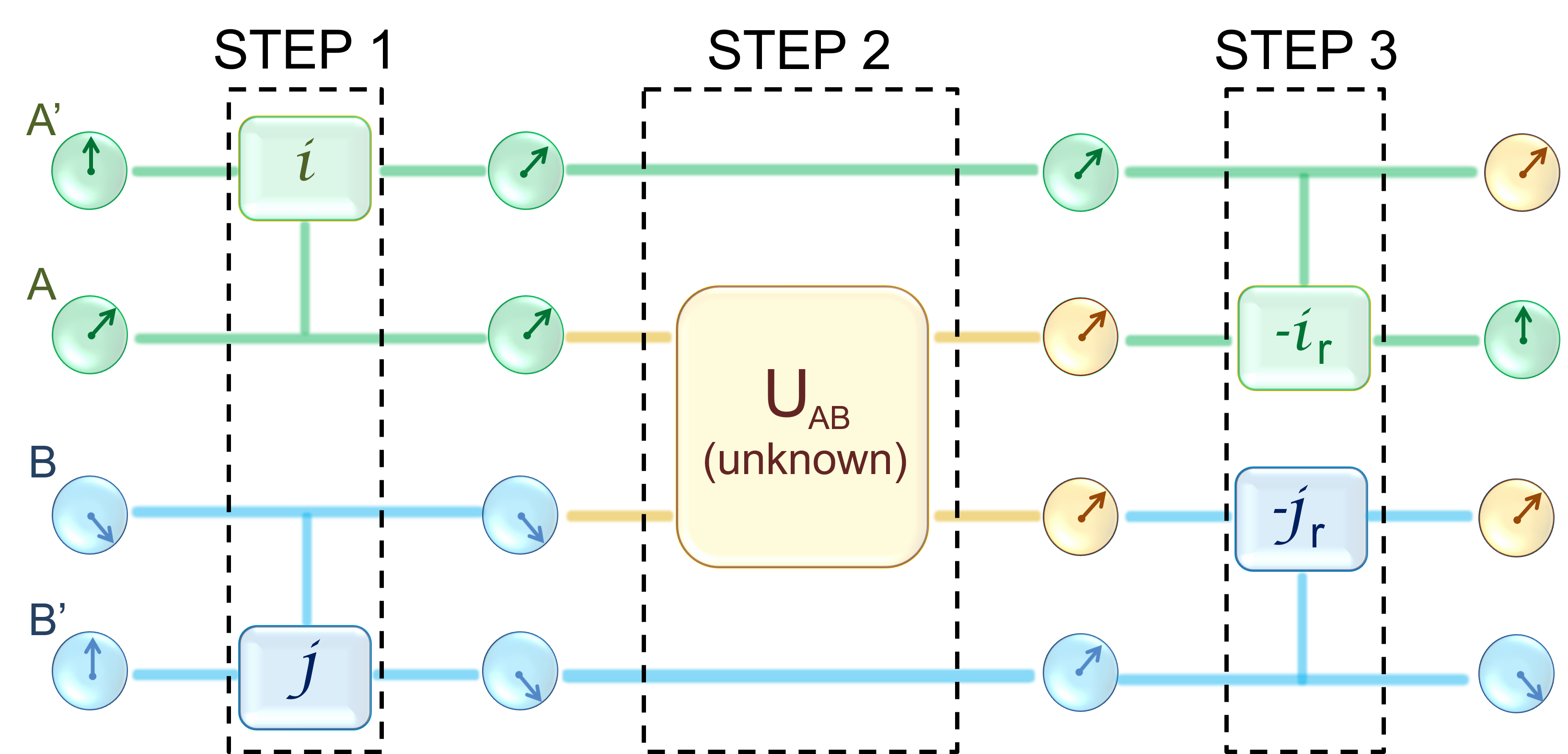}
\caption{Quantifying OWI.  The causal influence ${\cal C}_U(A\rightarrow B)$ exerted by $A$ on $B$ during the joint evolution  $U_{AB}$ is quantified by implementing the depicted scheme, STEP 1-3, and computing the measure defined in Eq.~\ref{def}. Specifically, correlations in the final state inform about a OWI flow generated by $U$.}
\label{fig1}
\end{figure}
 
\noindent Yet, there is a way to discern OWI from the initial and final states of a quantum process.  One can recast the problem of inferring causation in terms of the much better understood task of quantifying correlations, if additional systems are available. I first discuss an illustrative example. Then, I detail a generally applicable scheme.\\ 
Suppose one correlates two systems $A,B$ with two auxiliary systems $A',B'$, respectively, such that the global (pure) state is $\psi_{A'ABB'}:=\sum_{ij} c_i d_j\ket{i i}_{A'A} \ket{jj}_{BB'}$.  Consider then three different processes: 
\begin{eqnarray}\label{eq1}
&&V_{A'A} W_{BB'} U^1_{AB}\ \ \psi_{A'ABB'}= \sum_{ij} c_i d_j \ket{i}_{A'} \ket{0}_A  \ket{0}_{B} \ket{j}_{B'},\\
&&V_{A'A} W_{BB'}U^2_{AB}  \ \ \psi_{A'ABB'}=\sum_{ij} c_i d_j   \ket{i}_{A'}\ket{0}_{A} \ket{i}_{B}\ket{j}_{B'} , \nonumber\\
&&V_{A'A} W_{BB'}U^3_{AB}\ \ \psi_{A'ABB'}=\sum_{ij} c_i d_j  \ket{i}_{A'} \ket{j}_{A} \ket{0}_{B}\ket{j}_{B'}, \nonumber\\
&&V_{A'A} W_{BB'}=C^{j\rightarrow -j}_{B'\rightarrow B}C^{i\rightarrow -i}_{A'\rightarrow A},  U^1_{AB}=I_{AB}, U^2_{AB}= C^{i\rightarrow i}_{A\rightarrow B},\nonumber\\
   &&U^3_{AB}=C^{j\rightarrow j}_{B\rightarrow A}.\nonumber
\end{eqnarray} 
\noindent In the first case, there is no interaction between $A$ and $B$,  since $U^1_{AB}=I_{AB}$. The two final  controlled operations $V,W$ destroy all the initial correlations. In the second line, instead, a controlled gate  $U^2_{AB}$ generates four-partite correlations by sending information from $A$ to $B$. The subsequent controlled gates $V,W$ leave the systems $A',B$ correlated. In the third case, $U^3_{AB}$ generates a reverse information flow,  and correlations between systems $A,B'$ survive in the output state. 
Hence, the direction of the information flow between $A$ and $B$, if any, is determined from the correlations in $A'ABB'$.    \\

\noindent The example suggests a universally valid  scheme for quantifying OWI between two $d$-dimensional   systems $A,B$, due to an unknown channel $U_{AB}$, with respect to   reference bases $\{i_r\},\{j_r\}$  (Fig.~\ref{fig1}):\\
STEP 1 --  Given the initial state $\rho_A\otimes \rho_B$, with $\rho_A=\sum_{ik}\rho_{ik} \ket{i}\bra{k}_{A}$ and  $\rho_B=\sum_{jl}\rho_{jl} \ket{j}\bra{l}_{B}$, apply two controlled operations that create correlations between $A,B$ and two additional systems $A',B'$.  Defining $C^{j\rightarrow j; i\rightarrow i}_{B\rightarrow B'; A\rightarrow A'} :=C^{j\rightarrow j}_{B\rightarrow B'}C^{i\rightarrow i}_{A\rightarrow A'}$, one has
\begin{eqnarray}
&& \rho^{{\text in}}_{A'ABB'}:=C^{j\rightarrow j; i\rightarrow i}_{B\rightarrow B'; A\rightarrow A'}\ket{0}\bra{0}_{A'}\otimes\rho_{A}\otimes \rho_{B}\otimes\ket{0}\bra{0}_{B'} C^{l\rightarrow l; k\rightarrow k}_{B\rightarrow B'; A\rightarrow A'},\nonumber\\
&& \rho^{{\text in}}_{A'A}=\sum_{ik} \rho_{ik}\ket{ii}\bra{kk}_{A'A}, \ \ \ \rho^{{\text in}}_{BB'}=\sum_{jl} \rho_{jl} \ket{jj}\bra{ll}_{BB'}.
 \end{eqnarray}
    The bases $\{i\},\{j\}$  are from now on fixed to be  mutually unbiased with respect to the  eigenbases of $\rho_A, \rho_B$, while the method works for any choice but the  eigenbases themselves.  
  For instance, for pure input states, the basis $\{i\}$ is one in which the  input  is maximally coherent, $\rho_{ik}=e^{i\theta_{ik}}/d, \forall i,k, \theta$, such that the controlled operation creates maximal entanglement with the related auxiliary, $\sum_{ik} e^{i\theta_{ik}}\ket{ii}\bra{kk}/ d$, as the initial marginal states $\text{Tr}_{A'A}\psi_{A'ABB'}, \text{Tr}_{BB'}\psi_{A'ABB'}$ of the example. For maximally mixed input states, $\rho_{ik}=\delta_{ik}/d$, one obtains a maximally correlated classical state, $\sum_{i}\ket{ii}\bra{ii}/d$. 
  
 \noindent STEP 2 -- Let the  system $AB$  evolve according to the   channel $U $, 
 \begin{eqnarray}
 \rho_{A'ABB'}^{\text{in},U}:= \big(I_{A'}\otimes U_{AB} \otimes I_{B'}\big) \rho^{{\text in}}_{A'ABB'} \big(I_{A'}\otimes U_{AB}^{\dagger} \otimes I_{B'}\big).
  \end{eqnarray}
 
 \noindent STEP 3 (final) -- Apply a  second pair of local controlled operations with respect to the reference bases $\{i_r\},\{j_r\}$, but  swapping the roles of control and target systems:
  \begin{eqnarray}\label{forproof}
\rho_{A'ABB'}^{\text{f}}:=C^{j_r \rightarrow - j_r; i_r\rightarrow - i_r}_{B'\rightarrow B; A'\rightarrow A }    \rho^{\text{in},U}_{A'ABB'}C^{j_r \rightarrow - j_r; i_r\rightarrow - i_r}_{B'\rightarrow B; A'\rightarrow A }.
 \end{eqnarray}

\noindent The case study in Eq.~\ref{eq1} implies that one can evaluate the information exchanged by $A$ and $B$ by calculating the correlations  in  $\rho_{A'ABB'}^{\text{f}}$. The statistical dependence between   two systems $\alpha, \beta$ is quantified by the mutual information ${\cal I}(\alpha:\beta):=S(\alpha)+ S(\beta)-S(\alpha\beta),$ in which   $S(\alpha):=-\text{Tr}\{\alpha\log_2 \alpha\}$  is the von Neumann entropy of the state $\rho_\alpha$.
For any third system $\gamma$, the conditional mutual information reads  ${\cal I}(\alpha:\gamma|\beta):=  {\cal I}(\alpha:\beta\gamma)-{\cal I}(\alpha:\beta)$ 
\cite{cover}.
  I propose to  measure
the OWI that $A$ sends to $B$ via the channel $U$ by
\begin{eqnarray}\label{def}
{\cal C}_U(A\rightarrow B): ={\cal I}(B:A'A|B'),
 \end{eqnarray}
which is computed on $\rho_{A'ABB'}^{\text{f}}$ \cite{notanome}.
 Consequently, the influence of $B$  on $A$ during the interaction under study is given by ${\cal C}_U(B\rightarrow A)= {\cal I}(A:BB'|A')$.\\ 
  As a minimal working example, consider two qubits in the input   state   $\ket{+0}_{AB}$,  and the  unitary map to be the CNOT gate $C^{0,1\rightarrow 0,1}_{A\rightarrow B}$. By applying the proposed scheme, one has
  
 {\small
   \begin{eqnarray}
 && \ket{+0}_{AB}= \frac{(\ket{0}+\ket{1})_A  (\ket{+}+\ket{-})_B}{2} \xRightarrow{\text{STEP 1}} \nonumber \\
  &&\frac{(\ket{00} +\ket{11})_{A'A} (\ket{++} +\ket{--})_{BB'}}{2}=\nonumber \\
  &&\frac{(\ket{00} +\ket{11})_{A'A} (\ket{00} +\ket{11})_{BB'}}{2}\xRightarrow{\text{STEP 2}}\nonumber\\
&&C^{0,1\rightarrow 0,1}_{A\rightarrow B} \frac{(\ket{00} +\ket{11})_{A'A} (\ket{00} +\ket{11})_{BB'}}{2}=\nonumber\\
&&\frac{\{\ket{00}_{A'A} (\ket{00} +\ket{11})_{BB'}+ \ket{11}_{A'A} (\ket{10} +\ket{01})_{BB'}\}}{2}\nonumber
 \xRightarrow{\text{STEP 3}}\nonumber\\
&& C^{0,1 \rightarrow 0,1; 0,1\rightarrow 0,1}_{B'\rightarrow B; A'\rightarrow A} \frac{ \ket{00}_{A'A} (\ket{00} +\ket{11})_{BB'}+ \ket{11}_{A'A} (\ket{10} +\ket{01})_{BB'} }{2}\nonumber\\
&& = \frac{(\ket{0}   \ket{0}  \ket{0} +\ket{1}  \ket{0}  \ket{1})_{A'AB}}{\sqrt2}\ \ \ket{+}_{B'}.\nonumber
  \end{eqnarray}
  }
 \noindent As expected, the output state in the example displays correlations between $A'$ and $B$,   ${\cal I}(B:A'A|B')=2, {\cal I}(A:BB'|A')=0$.  Indeed, $A$ causally influences $B$.
 
 \section{PROOFS THAt the OWI measure satisfies desirable properties, including extension to the multipartite case}  
 
  \noindent   To further justify the proposal, I report other explicit calculations for instructive cases in Tables~\ref{tab1}, \ref{tab2}. Also, I discuss how the measure meets several desirable properties.\\

 \begin{table}[t]
  \bgroup
\def\arraystretch{1.5}
\begin{tabular}{|c|c|c| } 
\hline
 \textbf{\text{Process:}}\  $\bm{U_{AB}\ \rho_{AB}\ U_{AB}^{\dagger}}$ &    $\bm{{\cal C}_U(A\rightarrow B)}$&    $\bm{{\cal C}_U(B\rightarrow A)}$    \\
 \hline
$V_{A}\otimes W_{B}\  \rho_{AB}\ (V_{A}\otimes W_{B})^{\dagger}$   & $0 $  & $0 $   \\
\hline
$C^{0,1,\ldots,d\rightarrow 0,1,\ldots,d}_{A\rightarrow B}\ \psi_{A}\otimes\phi_{B}$   & $2 \log_2 d $& $0 $   \\
\hline
$C^{0,1,\ldots,d\rightarrow 0,1,\ldots,d}_{B\rightarrow A}\ \psi_{A}\otimes\phi_{B}$  &$0 $&$2 \log_2 d $\\
\hline
$C^{0,1\rightarrow +,-}_{A\rightarrow B}\ \psi_{A}\otimes\phi_{B}\ (d=2)$  & $0 $ & $0 $    \\
\hline
$C^{+,-\rightarrow +,-}_{A\rightarrow B}\ \psi_{A}\otimes\phi_{B}\  (d=2)$  & 0& $2  $   \\
\hline
 $C^{0,1,\ldots,d\rightarrow 0,1,\ldots,d}_{A\rightarrow B}\  \sum_{ij} 1/d^2 \ket{ij}\bra{ij}_{AB}$      &   $\log_2 d $ &0 \\       
\hline  
 $C^{0,1,\ldots,d\rightarrow 0,1,\ldots,d}_{B\rightarrow A}\  \sum_{ij} 1/d^2 \ket{ij}\bra{ij}_{AB}$      &   0& $\log_2 d$   \\   
\hline
$\text{SWAP}_{A,B}\ \psi_{A}\otimes\phi_{B}$   & $2 \log_2 d $&      $2 \log_2 d $ \\   
\hline
$\text{SWAP}_{A,B}\ \sum_{ij} 1/d^2 \ket{ij}\bra{ij}_{AB}$       &    $\log_2 d$  &    $\log_2 d$ \\   
\hline
\end{tabular}  
\egroup
\caption{Quantifying OWI (by the measure defined in Eq.~\ref{def})  generated via a channel $U_{AB}$, for $d$-dimensional systems $A,B$, with respect to the computational bases $\{i_r\}= \{j_r\}=0,1,\ldots,d$. Note $\rho,\psi,\phi$ indicate arbitrary mixed and pure input states.  Local unitaries do not generate OWI. Controlled  gates create maximal OWI flow. If one picks mutually unbiased bases, e.g. $C^{+,-\rightarrow +,-}_{A\rightarrow B}$ in a two-qubit case, the  information transfers from $B$ to $A$. Moreover, the measure correctly detects the two-way information transfer due to the SWAP gate, in both quantum and classical scenarios.} 
\label{tab1}
\end{table}

  \begin{table*}[t]
  
  {\footnotesize
  \bgroup
\def\arraystretch{1.5}
\begin{tabular}{|c|c|c|c|} 
\hline
 \textbf{Process:}\   $\bm{U_{EAB}\ \rho_{EAB}\ U_{EAB}^{\dagger}}$ &    $\bm{{\cal C}_U(EA\rightarrow B)}$ & $\bm{{\cal C}_U(A\rightarrow B|E)}$ & $\bm{{\cal C}_{U}(A\rightarrow B)}$   \\
 \hline
$V_{EA}\otimes W_{B}\  \rho_{EAB}\  (V_{EA}\otimes W_{B})^{\dagger}$   & $0 $   &$0 $&$0 $\\
\hline
$C^{0,1,\ldots,d\rightarrow 0,1,\ldots,d}_{A\rightarrow B}\  \psi_{EA}\otimes\phi_{B}$     & $2 \log_2 d $ & $\log_2 d $ & 
$\log_2 d $  \\
\hline
$C^{0,1,\ldots,d\rightarrow 0,1,\ldots,d}_{B\rightarrow A}\ \psi_{EA}\otimes\phi_{B}$    & $0 $&  $0 $ & $0 $\\
\hline
$C^{0,1,\ldots,d\rightarrow 0,1,\ldots,d}_{A\rightarrow B}\  \xi_E\otimes\psi_{A}\otimes\phi_{B}$      &     $ 2 \log_2 d $ & $ 2\log_2 d $& $ 2 \log_2 d $\\   
\hline
$C^{0,1,\ldots,d\rightarrow 0,1,\ldots,d}_{A\rightarrow B}\   \sum_{ij} 1/d^2 \ket{ii}\bra{ii}_{EA} \ket{j}\bra{j}_{B}$      &   $\log_2 d$    & $0$&$\log_2 d$\\   
\hline
$\text{CCNOT}_{EA\rightarrow B}\  (\ket{00}+\ket{11})_{EA}/\sqrt2\otimes\phi_{B}\  \   (d=2)$      &     $2 $ &$1 $ & $1 $\\ 
\hline  
$\text{CCNOT}_{EA\rightarrow B}\  \xi_E\otimes\psi_{A}\otimes\phi_{B}\ \  (d=2)$      &    $3/2 \log4/3+1 < 2 $ &$3/4 \log4/3+1/2 <1 $ &$3/4 \log4/3+1/2 $\\   
\hline
$\text{CCNOT}_{EA\rightarrow B}\   \sum_{ij} 1/4 \ket{ii}\bra{ii}_{EA} \ket{j}\bra{j}_{B}\ \ (d=2)$         &     $1 $ &$0$ & $1 $\\   
\hline
$\text{CCNOT}_{EA\rightarrow B}\  1/8\sum_{mij}  \ket{mij}\bra{mij}_{EAB}\ \ (d=2)$      &    $3/4 \log4/3+1/2$ &$1/2$ &$3/4 \log4/3$\\   
\hline
\end{tabular}
\egroup
}
\caption{Quantifying  OWI in tripartite classical and quantum  dynamics $U_{EAB}$ (by the quantities defined in Eqs.~\ref{def},\ref{res2}), including local unitaries, controlled gates, and two-qubit Toffoli gates (CCNOT), with respect to the reference bases $\{i_r\}, \{j_r\}, \{m_r\}=0,1,\ldots,d$. Note $\rho,\psi,\phi,\rho$ indicate arbitrary mixed and pure input states, and the $\text{CCNOT}_{EA\rightarrow B}$ gate transfers information from the control system $EA$ to the target $B$. The conditional causation ${\cal C}_U(A\rightarrow B|E)$  discriminates, for example, between a  controlled gate between $A$ and $B$ with $E$ uncorrelated, and one with correlated $EA$.  It is generalized to quantify OWI flow in multipartite systems of arbitrary size  by Eq.~\ref{res3}.}
\label{tab2}
\end{table*}

\begin{figure}[t!]
\centering
\includegraphics[width=.47\textwidth,height=4.5cm]{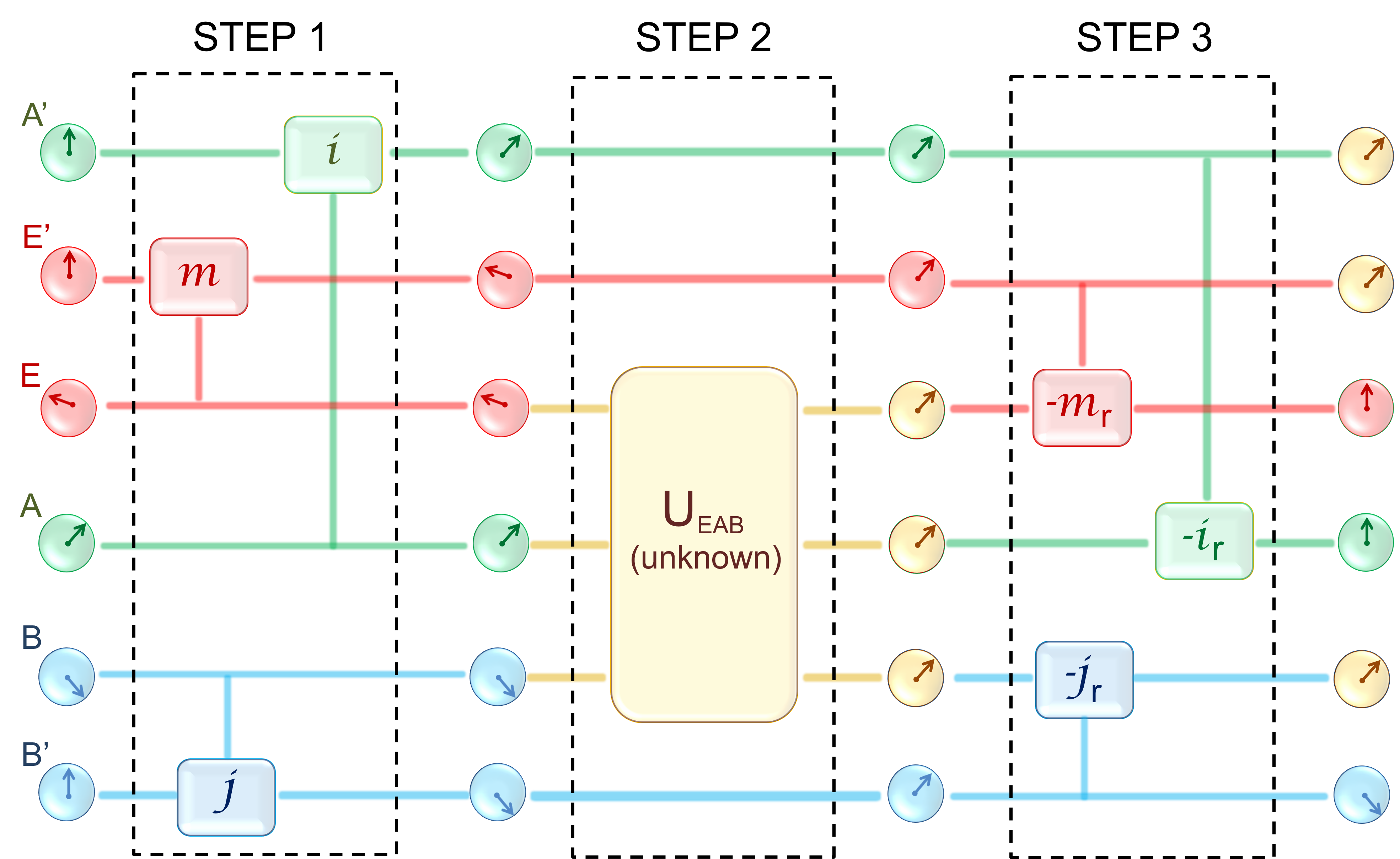}
\caption{Quantifying OWI in tripartite systems. The quantity ${\cal C}_{U}(A\rightarrow B|E)$ measures the control exerted   by $A$ on $B$ during the evolution $U_{EAB}$, given full information on $E$.}
\label{fig2}
\end{figure}

\noindent {\it Information-theoretic consistency}. There is no OWI without interaction. 
For local unitaries $U_{AB}=U_{A}\otimes U_{B}$,
one has 
${\cal C}_{U_A\otimes U_{B}}(A(B)\rightarrow B(A))=0$. Two systems can influence each other by a  two-way information flow, e.g. $V_{AB}\ket{+}_A\ket{+}_B= \ket{+}_A\ket{+}_B$, with $V_{AB}= C^{0,1\rightarrow 0,1; 0,1\rightarrow 0,1}_{A\rightarrow B; B\rightarrow A}$. In such a case, ${\cal C}_{V_{AB}}(A\rightarrow B)={\cal C}_{V_{AB}}(B\rightarrow A)=2$. Yet, the measure is not additive.  Given $U_{AB}=V_{AB}W_{AB}$, in general ${\cal C}_{U}(A\rightarrow B)\neq {\cal C}_{V}(A\rightarrow B)+{\cal C}_{W}(A\rightarrow B)$. Indeed, a controlled operation with control $A$ and target $B$ can be transformed in one with control $B$ and target $A$ by local unitaries, e.g. $C^{0,1\rightarrow 0,1}_{A\rightarrow B}= H_A\otimes H_B C^{0,1\rightarrow 0,1}_{B\rightarrow A} H_A\otimes H_B$, where $H$ is the Hadamard gate.\\ The measure is maximized by a controlled operation with respect to the reference bases and pure input states, ${\cal C}_U(A\rightarrow B) = 2\log_2 d$.  The unitary  creates $\log_2 d$ bits of classical correlations between $A'A$ and $BB'$, and $\log_2 d$ bits of quantum correlations, which are generated by consuming local coherence with respect to the local $A'A$ basis $\{i\}\otimes\{i\}$ \cite{jajun}.  For a maximally mixed input state, one has ${\cal C}_U(A\rightarrow B)=\log_2 d$, because only $\log_2 d$ bits of classical correlations are created.\\ 
\noindent {\it Asymmetry}. The measure, unlike correlation quantifiers, captures the direction of OWI,  ${\cal C}_U(A\rightarrow B)\neq {\cal C}_U(B \rightarrow A)$. Consider  $C^{0,1\rightarrow 0,1}_{A\rightarrow B}\ket{+0}_{AB}=(\ket{00}+\ket{11})_{AB}/\sqrt2$. Evaluating OWI with respect to $i_r=j_r=\{0,1\}$, one has ${\cal C}_{C^{0,1\rightarrow 0,1}_{A\rightarrow B}}(A\rightarrow B)=2,$ and ${\cal C}_{C^{0,1\rightarrow 0,1}_{A\rightarrow B}}(B\rightarrow A)=0$. On the other hand, reminding that $C^{0,1\rightarrow 0,1}_{A\rightarrow B}=C^{+,-\rightarrow +,-}_{B\rightarrow A}$,  the OWI with respect to $i_r=j_r=\{+,-\}$ is ${\cal C}_{C^{0,1\rightarrow 0,1}_{A\rightarrow B}}(A\rightarrow B)=0,$ and ${\cal C}_{C^{0,1\rightarrow 0,1}_{A\rightarrow B}}(B\rightarrow A)=2$. The measure correctly identifies control (the information source) and target (the affected system).  \\
 \noindent {\it Quantifying OWI with and without correlations.} One of the main challenges in evaluating  OWI is discriminating causal links between correlated systems.  
 The measure defined in Eq.~\ref{def} takes zero value for systems $A,B$ that do not exchange information, regardless of the presence of correlations. That is, two correlated systems $AB$ are left correlated by local unitary channels, but there is no information flow.
  A technical {\it caveat}  is that in the detection scheme the initial correlations between $A$ and $B$ must be ignored. The input state is  $\rho_{A}\otimes \rho_B$, rather than the full state $\rho_{AB}$. \\ A more elusive manifestation of OWI is when there is influence without correlations, e.g. $C^{0,1\rightarrow 0,1}_{A\rightarrow B}\ket{10}_{AB}=\ket{11}_{AB}$.   The measure is able to detect  such causal relations,  discriminating  when  $U$ is a controlled operation and when  the very same input/output transformation is due to a local unitary, $X_B \ket{10}_{AB} =\ket{11}_{AB} $. Note that OWI flow is detected even with no state change, e.g. $C^{0,1\rightarrow 0,1}_{A\rightarrow B}\ket{00}_{AB}=\ket{00}_{AB}$, as the system $B$ still receives the instruction ``do nothing'' from $A$. Indeed, the controlled gate, while generating no change, is a distinct physical process from the identity channel.\\
 \noindent {\it Quantifying classical OWI.} A measure of OWI transfer between random variables $A,B$, which overcomes some limitations of previous proposals,  is obtained by considering classically correlated input states $\sum_{kl} p_{kl} \ket{kl}\bra{kl}_{AB}$ \cite{piani}. 
The celebrated Granger causality \cite{granger}, a causation measure widely used in econometrics, is confined to detect linear causal relations. On the same hand, the  transfer entropy  and the   causation entropy \cite{infotransfer,causal},  employed in network science, fail to detect causation generated by logic gates, such as the CNOT (XOR) applied to time series,  $A_t=A_{t-1}, B_t=A_{t-1}\oplus B_{t-1}$, and  the SWAP, $A_t=B_{t-1}, B_{t}=A_{t-1}$ \cite{james,janzing}.  The quantity in Eq.~\ref{def} instead correctly describes causal relations implemented by classical  gates (see Table \ref{tab1}). For instance, for maximally mixed inputs  one has ${\cal C}_{\text{SWAP}_{A,B}}(A\rightarrow B)={\cal C}_{\text{SWAP}_{A,B}}(
B\rightarrow A)=\log_2 d$. The result is independent of the chosen decomposition for the SWAP gate in terms of controlled gates \cite{swap}.
 Moreover, a  surprising result is obtained: Quantum devices can detect OWI between {\it classical} systems that is untraceable by  classical machines, if the proposed evaluation scheme is adopted. Consider the process $C^{0,1\rightarrow 0,1}_{A\rightarrow B}  \ket{0}\bra{0}_A\ket{0}\bra{0}_B=\ket{0}\bra{0}_A\ket{0}\bra{0}_B$,  in which two systems $A,B$ carry information about two  random variables.  
   If $A'ABB'$ is a classical four-bit register, no superpositions with respect to the basis $\{0,1\}^{\times 4}$ are possible. Applying  the method in Fig.~\ref{fig1}, the final state is $\ket{0000}\bra{0000}_{A'ABB'}$.  Hence, no OWI can be detected by classical means.  If instead we store the information about the classical variables into the states   of two qubits $A,B$, and the auxiliary $A',B'$ are quantum systems as well,  quantum correlations are created by STEP 1-3. One then obtain the same final state of the working example,  $1/\sqrt2\ \  (\ket{0}   \ket{0}  \ket{0} +\ket{1}  \ket{0}  \ket{1})_{A'AB}\ \ \ket{+}_{B'}$. Since  classical variables are under scrutiny,  an additional  STEP 4 is included to offset the (fictitious) quantum correlations: Projecting into the classical basis $\{0,1\}^{\times 4}$,  one has
 \begin{eqnarray}
 \frac{\ket{000}\bra{000}+\ket{101}\bra{101}_{A'AB}}{2} \otimes I_{B'}/2.\nonumber
 \end{eqnarray}
Computing the measure defined in Eq.~\ref{def} on the projected state gives ${\cal C}_{C_{A\rightarrow B}^{0,1\rightarrow 0,1}}(A\rightarrow B)=1$. The result is expected as one bit of correlations is created between  $A'A$ and $BB'$. \\
{\it Scalability: Localizing and ranking multiple, concurrent  information sources.}  The proposed OWI measure  extends  to  many-body systems. 
 Defining a third system $E$, the global evolution of the tripartition    in the state $\rho_{EAB}$ is  a unitary $U_{EAB}$. A generalization of the OWI detection scheme, as depicted in Fig.~\ref{fig2}, allows for quantifying the  influence of, say, system $A$ over $B$ in the presence of $E$, reconstructing the causal pattern between the three systems. Alike $A,B$, the system $E$ is coupled with an auxiliary  $E'$, 
  \begin{eqnarray}
  C^{m\rightarrow m}_{E\rightarrow E'}  \ket{0}\bra{0}_{E'} \sum\limits_{mn} \rho_{mn} \ket{m}\bra{n}_{E}C^{n\rightarrow n}_{E\rightarrow E'}= \sum\limits_{mn} \rho_{mn} \ket{mm}\bra{nn}_{E'E}.\nonumber
  \end{eqnarray} 
 One then obtains the sixpartite state {\small $\rho^{{\text in}}_{E'A'EABB'}$} (STEP 1), the evolved state $\rho^{{\text in},U}_{E'A'EABB'}=U_{EAB}\rho^{{\text in}}_{E'A'EABB'}U_{EAB}^{\dagger}$ (STEP 2), and the final state (STEP 3)
\begin{eqnarray}
 \rho^{{\text f}}_{E'A'EABB'}:= C^{m_r\rightarrow -m_r,j_r\rightarrow -j_r, i_r\rightarrow -i_r}_{E'\rightarrow E,B'\rightarrow B, A'\rightarrow A}\rho^{{\text in},U}_{E'A'EABB'}C^{m_r\rightarrow -m_r,j_r\rightarrow -j_r, i_r\rightarrow -i_r}_{E'\rightarrow E,B'\rightarrow B, A'\rightarrow A.}\nonumber
\end{eqnarray}
\noindent The degree of control of $A$ on $B$ {\it given full information about  $E$}, with respect to the reference bases $\{i_r\},\{j_r\},\{m_r\}$, is then quantified by the difference between the OWI from  $AE$ to $B$  and the OWI from $E$ alone  ($A$ is ignored),
\begin{eqnarray}\label{res2}
{\cal C}_U(A\rightarrow B|E):&=& 
{\cal C}_U(EA\rightarrow B)-{\cal C}_U(E\rightarrow B)\\
&=&{\cal I}(B: A'A|E'EB'),\nonumber
\end{eqnarray}
 computed in the final state $\rho^{{\text f}}_{E'A'EABB'}$. 
The quantity, while giving  different results from the OWI evaluated without information on $E$, ${\cal C}_{U_{EAB}}(A\rightarrow B)\neq{\cal C}_{U_{EAB}}(A\rightarrow B|E)$, inherits by construction the consistency and asymmetry properties. 
 It is also explicitly computable for tripartite dynamics of classical and quantum  systems, including Toffoli and bipartite controlled gates (Table \ref{tab2}). A quantifier of classical conditional causation as a special case is obtained by computing the conditional mutual information in the final state after it is projected into the reference bases. 
   The extension to many-body processes of arbitrary size is straightforward. Given an $N$-partite system $\cup_{\alpha=1}^NS_\alpha$  evolving via the unitary  $U_{1\ldots N}$,  the OWI sent from a subsystem $S_\alpha$ to a subsystem $S_\beta$ is  
   \begin{eqnarray}\label{res3}
   {\cal C}_{U_{1\ldots N}}(S_\alpha\rightarrow S_\beta| \mathsmaller{\bigcup\limits_{\gamma\neq\alpha,\beta}} S_\gamma)={\cal I}(S_\beta: S_\alpha'S_\alpha|S_\beta' \mathsmaller{\bigcup\limits_{\gamma\neq\alpha,\beta}}S_\gamma'S_{\gamma}). 
   \end{eqnarray}
   The chain rule of the conditional mutual information   implies that the  total OWI received from  a subsystem is  decomposable as the sum of conditional causations.  \\

 \section{Conclusion} 
   I have introduced a scheme  to evaluate OWI (one-way information) generated via a quantum channel  (Fig.~\ref{fig1}). Then, I have built an  information-theoretic measure of OWI, Eq.~\ref{def}. 
  The study paves the way for a resource theory of  OWI \cite{gour},   a mathematical framework studying   the computational power of causal, one-way information flow \cite{costajin}. OWI, rather than correlation, could be the key resource when different parts of a composite system play different roles, e.g. control \cite{control}, metrology \cite{metrology}, and learning \cite{learning}. As  OWI can be evaluated from correlation dynamics, one may  build measures of genuine quantum and classical information flow, as it happens for correlation quantifiers \cite{modirev}.  
 
\section{Acknowledgements}
I thank Chao Zhang for useful comments.  The research presented in this  article was supported by the R. Levi Montalcini Fellowship of the Italian Ministry of Research and Education, and the Laboratory Directed Research and Development program of Los Alamos National Laboratory under project number 20180702PRD1. Los Alamos National Laboratory is managed by Triad National Security, LLC, for the National Nuclear Security Administration of the U.S.
Department of Energy under Contract No. 89233218CNA000001.


\begin{thebibliography}{99}
\bibitem{note}No correlations in both input and output states does not preclude interaction. A physical, continuous time implementation of a controlled gate $U_{AB,t}$ from an uncorrelated input to an uncorrelated output may generate correlations at some time $t$.
\bibitem{notanome} The letter ${\cal C}$ is picked as the measure quantifies the strength of the causal link or, in other words, the degree of Control of a system on another.
\end{thebibliography}
\end{document}